\documentclass[a4paper,12pt]{article}
\pdfoutput=1
\usepackage{feynmp-auto,expdlist}
\usepackage{amsmath, amsfonts, amssymb,mathbbol}
\usepackage{graphicx}
\usepackage{enumerate,cancel}
\usepackage{hyperref}
\usepackage{latexsym}
\usepackage{dsfont}
\usepackage{hepnicenames}
\usepackage{enumerate}
\usepackage{soul}
\usepackage[normalem]{ulem}
\usepackage{comment}

\usepackage{units}

\newcommand{\gfootnote}[1]{\footnote{\tiny\color{gray}#1}}
\renewcommand{\gfootnote}[1]{}
\usepackage{soul}

\usepackage{mathrsfs,graphicx,rotating,amsmath,amsfonts,mathtools,booktabs,amssymb,wasysym}
\usepackage{hyperref}\usepackage{slashed}
\usepackage[nosort]{cite}
\usepackage[table,xcdraw,dvipsnames]{xcolor}
\usepackage{bm}
\usepackage{graphicx}
\usepackage{multirow,multicol}
\hypersetup{colorlinks,bookmarksopen,bookmarksnumbered,
	linkcolor=blus,pdfstartview=FitH,urlcolor=rossos,citecolor=verde}
\allowdisplaybreaks

\renewcommand{\[}{\left[}

\newcommand{\Lag}{\mathscr{L}}

\newcommand{\mio}[1]{}

\newcommand{\med}[1]{\langle #1\rangle}

\newcommand{\bpm}{\begin{pmatrix}}
\newcommand{\epm}{\end{pmatrix}}

\usepackage{mathrsfs}

\newcommand{\fig}[1]{~\ref{fig:#1}}

\renewcommand{\Im}{{\rm Im}\,}
\renewcommand{\Re}{{\rm Re}\,}
\allowdisplaybreaks
\usepackage{multicol}
\usepackage{color}
\definecolor{rosso}{cmyk}{0,1,1,0.4}
\definecolor{rossos}{cmyk}{0,1,1,0.55}
\definecolor{rossoc}{cmyk}{0,1,1,0.2}
\definecolor{blu}{cmyk}{1,1,0,0.3}
\definecolor{blus}{cmyk}{1,1,0,0.6}
\definecolor{bluc}{cmyk}{1,1,0,0.1}
\definecolor{verde}{cmyk}{0.92,0,0.59,0.25}
\definecolor{verdec}{cmyk}{0.92,0,0.59,0.15}
\definecolor{verdes}{cmyk}{0.92,0,0.59,0.4}

\newcommand{\bp}{\bar{M}_{\rm Pl}}
\newcommand{\eq}[1]{~{\rm (\ref{eq:#1})}}

\newcommand{\GeV}{\,{\rm GeV}}
\newcommand{\TeV}{\,{\rm TeV}}

\newcommand{\beq}{\begin{equation}}
\newcommand{\eeq}{\end{equation}}

\newcommand{\bea}{\begin{eqnarray}}
\newcommand{\eea}{\end{eqnarray}}
\newcommand{\be}{\begin{equation}}
\newcommand{\ee}{\end{equation}}
\font\tenrsfs=rsfs10 at 12pt
\font\sevenrsfs=rsfs7
\font\fiversfs=rsfs5
\newfam\rsfsfam
\textfont\rsfsfam=\tenrsfs
\scriptfont\rsfsfam=\sevenrsfs
\scriptscriptfont\rsfsfam=\fiversfs
\def\be#1\ee{\begin{equation}#1\end{equation}}
\def\bl#1\el{\begin{align}#1\end{align}}
\def\ba#1\ea{\begin{align*}#1\end{align*}}

\renewenvironment{thebibliography}[1]
{\begin{multicols}{2}[\section*{\refname}]%
		\@mkboth{\MakeUppercase\refname}{\MakeUppercase\refname}%
		\list{\@biblabel{\@arabic\c@enumiv}}%
		{\settowidth\labelwidth{\@biblabel{#1}}%
			\leftmargin\labelwidth
			\advance\leftmargin\labelsep
			\@openbib@code
			\usecounter{enumiv}%
			\let\p@enumiv\@empty
			\renewcommand\theenumiv{\@arabic\c@enumiv}}%
		\sloppy
		\clubpenalty4000
		\@clubpenalty \clubpenalty
		\widowpenalty4000%
		\sfcode`\.\@m}
	{\def\@noitemerr
		{\@latex@warning{Empty `thebibliography' environment}}%
		\endlist\end{multicols}}

\newcommand{\eV}{\,{\rm eV}}
\newcommand{\SU}{\,{\rm SU}}
\newcommand{\SL}{\,{\rm SL}}

\newcommand{\U}{\,{\rm U}}

\makeatletter
\font\ital=cmu10

\def\hhref#1{\href{http://arxiv.org/abs/#1}{arXiv:#1}}
\usepackage{xstring}
\newcommand{\hhrefq}[1]{\IfSubStr{#1}{:}{\href{http://inspirehep.net/search?ln=en&ln=en&p=#1&of=hb&action_search=Search&sf=&so=d&rm=&rg=25&sc=0}{InSpire:#1}}{\hhref{#1}}}

\def\art{\@ifnextchar[{\eart}{\oart}}
\def\eart[#1]#2#3#4#5#6{{\rm #2}, {\em #3 \bf #4} {\rm (#6) #5} ({\em #1})}
\def\article{\@ifnextchar[{\earticle}{\oarticle}}
\def\oarticle#1#2#3#4#5#6{{\rm #1}, {\ital `#6'}, {\rm #2 #3 (#5) #4}}
\def\earticle[#1]#2#3#4#5#6#7{{\rm #2}, {\ital `#7'}, {\rm #3 #4 (#6) #5}  [\hhrefq{#1}]}
\def\hepart[#1]#2{{\rm #2, \sl#1}}
\def\heparticle[#1]#2#3{#2, {\ital `#3'} [\hhrefq{#1}]}
\newcommand{\doi}[1]{\href{http://dx.doi.org/#1}{[link]}}

\newcommand{\hhrefqq}[1]{\IfBeginWith{#1}{10.}{\href{https://doi.org/#1}{doi:#1}}{\hhrefq{#1}}}

\renewenvironment{thebibliography}[1]
{\begin{multicols}{2}[\section*{\refname}]%
		\@mkboth{\MakeUppercase\refname}{\MakeUppercase\refname}%
		\list{\@biblabel{\@arabic\c@enumiv}}%
		{\settowidth\labelwidth{\@biblabel{#1}}%
			\leftmargin\labelwidth
			\advance\leftmargin\labelsep
			\@openbib@code
			\usecounter{enumiv}%
			\let\p@enumiv\@empty
			\renewcommand\theenumiv{\@arabic\c@enumiv}}%
		\sloppy
		\clubpenalty4000
		\@clubpenalty \clubpenalty
		\widowpenalty4000%
		\sfcode`\.\@m}
	{\renewcommand{\@noitemerr}
		{\@latex@warning{Empty `thebibliography' environment}}%
		\endlist\end{multicols}}

\newcommand{\eqnsystem}[1]{
	\renewcommand{\@eqnnum}{{\rm (\thealphaequation)}}
	\renewcommand{\@@eqncr}{\let\@tempa\relax \ifcase\@eqcnt \def\@tempa{& & &} \or
		\newcommand{\@tempa}{& &}\or \newcommand{\@tempa}{&}\fi\@tempa
		\if@eqnsw\@eqnnum\refstepcounter{alphaequation}\fi
		\global\@eqnswtrue\global\@eqcnt=0\cr}
	\refstepcounter{equation} \let\@currentlabel\theequation \def\@tempb{#1}
	\ifx\@tempb\empty\else\label{#1}\fi
	\refstepcounter{alphaequation}
	\let\@currentlabel\thealphaequation
	\global\@eqnswtrue\global\@eqcnt=0 \tabskip\@centering\let\\=\@eqncr
	$$\halign to \displaywidth\bgroup \@eqnsel\hskip\@centering
	$\displaystyle\tabskip\z@{##}$&\global\@eqcnt\@ne
	\hskip2\arraycolsep\hfil${##}$\hfil& \global\@eqcnt\tw@\hskip2\arraycolsep
	$\displaystyle\tabskip\z@{##}$\hfil
	\tabskip\@centering&\llap{##}\tabskip\z@\cr}
\def\endeqnsystem{\@@eqncr\egroup$$\global\@ignoretrue} \makeatother

\newcounter{alphaequation}[equation]
\renewcommand{\thealphaequation}{\theequation\hbox to
	0.6em{\hfil\alph{alphaequation}\hfil}}

\definecolor{Gray}{gray}{0.95}

\usepackage{tocloft}
\begin{document}
\begin{center}  
{\LARGE\bf\color{rossos} Baryogenesis from cosmological CP breaking} \\
\vspace{0.5cm}
 {\bf Mateusz Duch,$^a$} {\bf Alessandro Strumia$^a$} and {\bf Arsenii Titov$^{b}$}  \\[3mm]
{$^a$ \it Dipartimento di Fisica, Universit\`a di Pisa, Italia}\\
{$^b$ \it Dipartimento di Fisica e Astronomia, Universit\`a di Padova and INFN, Italia}\\
%{$^c$ \it INFN, Sezione di Padova, Italia}\\[1mm]

\vspace{0.7cm}
{\large\bf Abstract}
\begin{quote}{\large
We show that baryogenesis can arise from the cosmological evolution of a scalar field that governs CP-violating parameters, 
such as the Yukawa couplings and the theta terms of the Standard Model.
During the big bang, this scalar may reach a CP-violating minimum, where its mass can be comparable to the inflationary Hubble scale. 
Such dynamics can emerge in theories featuring either a spontaneously broken local U(1) symmetry or modular invariance. 
The latter arises naturally as the effective field theory capturing the geometric origin of CP violation in toroidal string compactifications. 
Modular baryogenesis is compatible with the  modular approach to resolving the strong CP problem.}
\end{quote}
\end{center}
\vspace{0cm}
\tableofcontents
\thispagestyle{empty}
\newpage

\section{Introduction}
A relativistic quantum field theory violates the CP symmetry when its physical parameters are complex. 
Sakharov identified CP violation as a crucial ingredient for generating the observed baryon asymmetry between particles and antiparticles, 
quantified as $n_B/n_\gamma =  (6.15\pm0.15)~10^{-10}$~\cite{2401.15054}.
In the Standard Model (SM), CP is violated %from
by the CKM phase, but its effects appear too small to account for the observed baryon asymmetry. 
 This suggests that the SM is the low-energy manifestation of a deeper, more fundamental theory. 
 Here, we explore the compelling possibility that this fundamental theory is real --- meaning it contains no complex parameters.

\smallskip

In section~\ref{sec:th}, we summarise theories where CP is spontaneously broken.
If the deeper real theory is another QFT, CP violation can occur through complex vacuum expectation values of scalar fields, denoted by $\tau$.
The deeper theory might be super-string theory.
Then  chiral fermions, the CP symmetry and its breaking can arise via compactification on a suitable space with a complex structure.
A flat torus is the simplest possibility.
As a remnant of this mechanism, the resulting effective QFT inherits a distinctive stringy modular symmetry
with a special modulus scalar  $\tau$ whose vacuum expectation value breaks CP.
Both possibilities result in an effective action where the phases of Yukawa couplings (including the CKM phase) 
and the theta angles non-trivially depend on $\tau$.

In section~\ref{sec:cosmo}, we show that the cosmological evolution $\tau(t)$ of the scalar, possibly towards a CP-breaking minimum
of its potential,
generically induces a baryon asymmetry, given that the $B$ and $L$ are broken by weak sphalerons and (likely) by Majorana neutrino masses. 
The time variation of the phase of the Yukawa couplings,
that naively just produces an irrelevant chiral asymmetry, actually leads to the baryon asymmetry given that 
SM interactions are chiral.

\smallskip

It is interesting to compare the CP-breaking scalar $\tau$ with the axion.
The axion was introduced as a solution to the strong CP problem, 
see~\cite{2003.01100} for a review.
More recently, it has been realised that $\tau$ provides an alternative solution to the strong CP problem~\cite{2305.08908,2404.08032,2406.01689}.
Some authors explored %``axiogenesis'':
the possibility that the cosmological evolution of a very light axion pseudo-scalar $a(t)$
around the weak phase transition contributes to baryogenesis~\cite{1407.0030,1811.03294}.
However, the resulting baryon asymmetry is too small to explain observations. 
Later, extensions that avoid this conclusion were proposed, coining the name ``axiogenesis''~\cite{1910.02080,2006.04809,2006.05687,2503.04888}.
The ``taugenesis'' scenario explored here differs from ``axiogenesis'' in a crucial way:
unlike the axion, the CP-breaking scalar $\tau$ lacks a shift symmetry because, by assumption, 
its vacuum expectation value directly affects physical parameters, such as the amount of CP violation.
No global U(1) symmetries are needed.
As a consequence, $\tau$ is much heavier and evolves at higher temperatures, during which weak sphalerons and Majorana neutrino masses significantly violate baryon and lepton number. This allows the generated baryon asymmetry to match the observed value.
Further comparisons are discussed in %the concluding 
section~\ref{sec:comparison}.

\section{Theories for spontaneous CP violation}\label{sec:th}
Theories with spontaneously broken CP symmetry must involve some complex structure. The most intuitive option is a U(1) symmetry, however in section~\ref{sec:U1gl}, we explain why a global U(1) is problematic. In section~\ref{sec:U1}, we explore how a local U(1) spontaneously broken by two scalars can break CP. Finally, in section~\ref{sec:mod}, we discuss the most plausible alternative: modular invariance. We begin by presenting the common effective field theory framework that describes these possibilities.

\begin{table}[t]
$$\displaystyle\begin{array}{|lc|ccc|cc|c|}\hline
\hbox{Fields}&{\rm spin} &{\rm U}(1)_Y&\SU(2)_L&\SU(3)_{\rm c}&L&B & d\cr \hline
U = u_R   &1/2& -{2 \over 3} & 1 & \bar{3} &\phantom{-}0&-\frac13 & 3\cr
D = d_R  &1/2& \phantom{-}{1 \over 3}& 1 &\bar{3} &\phantom{-}0&-\frac13 & 3\cr
Q=(u_L, d_L) &1/2 &\phantom{-} {1\over 6}  & 2 & 3&\phantom{-}0&\phantom{-}\frac13 & 6\cr 
E = e_R  &1/2&\phantom{-}1 & 1 &1  &-1&\phantom{-}0 & 1\cr
N=\nu_R &1/2& \phantom{-}0& 1 &1&-1&\phantom{-}0 & 1\cr
L=(\nu_L, e_L) &1/2& -{1 \over 2} & 2 &1&\phantom{-}1&\phantom{-}0 & 2\cr 
H = (0,v+h/\sqrt{2})& 0&\phantom{-}{1 \over 2} & 2 &1&\phantom{-}0&\phantom{-}0 & 2\cr  \hline
\end{array}$$
\caption{\em Quantum numbers of SM particles, including right-handed neutrinos $N$.
In the latter columns $L$ is lepton number, $B$ is baryon number, $d$ is the number of degrees of freedom.
\label{tab:SM}
}
\end{table}

\subsection{Effective field theory for spontaneous CP violation}\label{sec:eff}
We consider an effective QFT described by a Lagrangian that includes gauge-invariant kinetic terms for all particles, denoted as $\Lag_{\rm kin}$,
plus the usual non-minimal terms: Yukawa couplings and the scalar potential.
These couplings are assumed to be real functions of a CP-violating scalar field $\tau$, written as a dimensionless field.
To maintain invariance under $\tau$-dependent field redefinitions, the action must also include shift-invariant couplings of $\partial_\mu \tau$
to the number current $J_P^\mu$ of each particle as well as additional anomalous terms.
For example, in the SM only the Yukawa couplings can be complex,
and the matter content consists of the fields $P=\{Q,U,D,L,E,N, H\}$
as summarised in table~\ref{tab:SM}.
For later convenience and motivated by the observation of neutrino oscillations, we include right-handed neutrinos $N$.
The number current for the Higgs doublet $H$ is $J_H = i (H^\dagger \partial_\mu H - H \partial_\mu H^\dagger)$.
For Weyl fermions $\psi=\{Q,U,D,L,E,N\}$ the corresponding currents are $J_\psi = \bar \psi \bar\sigma^\mu \psi$.
Our assumptions lead to
\begin{eqnarray}
 \nonumber
 \Lag_{\rm eff} &=& \Lag_{\rm  kin} -
[  Y_{\rm u}(\tau) \,  QUH + Y_{\rm d}(\tau) QDH^*  + Y_{\rm e}(\tau) \,  LEH^* + Y_\nu(\tau) LNH+\hbox{h.c.} ] +\\
&& \label{eq:Lagtau}
%+
- \left[M(\tau) \frac{N^2}{2}+ \hbox{h.c.} \right] - V(H,\tau)+ \sum_P c_P(\tau) (\partial_\mu \tau)J^\mu_P +\\
&&
+\,\theta_3(\tau)   \frac{g_3^2}{32\pi^2} G^a_{\mu\nu} \tilde{G}^{a\mu\nu}+
\theta_2 (\tau) \frac{g_2^2}{32\pi^2} W^a_{\mu\nu} \tilde{W}^{a\mu\nu}+
\theta_1 (\tau)  \frac{g_Y^2}{16\pi^2} Y_{\mu\nu} \tilde{Y}^{\mu\nu} .\nonumber
\end{eqnarray}
Dual tensors $\tilde{F}=\{\tilde{G},\tilde{W},\tilde{Y}\}$ are defined as $\tilde{F}^{\mu\nu}=\frac{1}{2}\epsilon^{\mu\nu\rho\sigma}F_{\rho\sigma}$, with $\epsilon^{0123}=1$
so that $F_{\mu\nu}\tilde{F}^{\mu\nu}=-4\vec{E}\cdot\vec{B}$.
The fermionic currents $J^\mu_{Q,U,D,L,E}$ are violated by $\SU(3)_c\otimes\SU(2)_L\otimes\U(1)_Y$ anomalies
resulting in $\theta$ terms that induce (in the non-abelian case) weak and strong sphaleron processes.
As a result, the only conserved charge is hyper-charge, which, after the electroweak phase transition, becomes electric charge.
As is typical in the presence of anomalies, theoretical consistency requires accounting for both tree-level and loop-level effects.

\smallskip

Unlike in the axion case, our action is not invariant under $\tau\to \tau+\delta\tau$:
shift-invariance is broken by the Yukawa couplings $Y(\tau)$, such that $\tau$ controls the amount of CP violation.
The $Y(\tau)$, $c(\tau)$, and $\theta(\tau)$ functions of $\tau$
provide a redundant description of the same physics, as one can perform field redefinitions.
In particular, without loss of generality, one can remove all $c_P\to 0$ by performing a
$\tau$-dependent rephasing of each field $P$
\beq  \label{eq:rephasing}
P\to e^{i \phi_P(\tau) } P\qquad  \hbox{with}\qquad\phi_P(\tau)=\int^\tau d\tau\, c_P(\tau) .\eeq
This change of coordinates in field space affects the Yukawa couplings as 
\beq  \label{eq:Ytrans}
\begin{array}{ll}
Y_{\rm u}(\tau) \to e^{i (\phi_Q+\phi_U+ \phi_H) } Y_{\rm u}(\tau), ~~~~~& 
Y_{\rm d}(\tau) \to e^{i (\phi_Q+\phi_D- \phi_H) } Y_{\rm d}(\tau)\\
Y_{\rm e}(\tau) \to e^{i (\phi_L+\phi_E- \phi_H) } Y_{\rm e}(\tau), & 
Y_{\nu}(\tau) \to e^{i (\phi_L+\phi_N+ \phi_H) } Y_{\nu}(\tau)
\end{array}
\eeq
and the anomalous theta terms as~\cite{Fujikawa:1979ay}
\beq
\theta_3(\tau) \to \theta_3(\tau) - N_{\rm gen}  (2 \phi_Q+\phi_U+\phi_D),\qquad
\theta_2(\tau) \to \theta_2(\tau) - N_{\rm gen} (3 \phi_Q+ \phi_L).\eeq
%\AS{with $\gamma_5=\diag(-1,-1,+1,+1)$ 
For simplicity, we considered flavour-universal re-phasings.
In general $N_{\rm gen}=3$ is replaced by a sum over flavour indices.

\subsection{CP violation from spontaneously broken global U(1)?}\label{sec:U1gl}
As a first attempt of writing a theory where CP is a  spontaneously broken symmetry, 
let us consider  a {\em global} U(1), under which  particles $P=\{Q,U,H,\ldots\}$ carry global charges $q_P$.
This U(1) is spontaneously broken by a complex scalar $S = f e^{i \theta}/\sqrt{2}$.
Here and in what follows, we do not write the radial excitation for simplicity.
As a result, the theory contains a Goldstone boson, denoted as axion $a=\theta f$ rather than as $\tau$. 
The axion interacts via shift-invariant couplings, up to anomaly-induced terms.
Consequently, the Yukawa couplings can only depend on 
$a$ in a specific way. In the basis where there are no $\partial_\mu a$ couplings to particle currents, the Yukawas 
such as $Y_{\rm u}$ must have the form
\beq Y_{\rm u}(a)QUH = c S^{-q_{QUH}/q_S}  QU H \qquad \hbox{where}\qquad q_{QUH}\equiv q_Q+q_U + q_H \eeq 
where the coefficients $c$ are real constants that can be generated by integrating out additional heavy particles.
Rewriting the same action in a different form clarifies that the vacuum expectation value of $S$ does not break CP:
all particles can be rephased as $P\to e^{-i q_P a/q_S f}P$ rendering them neutral under the U(1) symmetry.
In this rephased basis, all couplings become  $a$-independent,
and kinetic terms of particles induce couplings to particle currents
\beq \sum_P \frac{q_P}{q_S} \frac{\partial_\mu a}{f} J^\mu_P. \eeq
In the $\sigma$-model formalism, this structure is described by a covariant-like derivative 
\beq \partial_\mu P \to D_\mu P  = \partial_\mu P - i \frac{q_P}{q_S}  \frac{\partial_\mu a}{ f}.\eeq
Breaking CP needs a more complicated structure: a U(1) broken by at least two scalars, such that their relative phase is physical.
However, global U(1) symmetries are generally expected to be accidental rather than fundamental, as they are typically broken by quantum gravity effects. This makes it challenging to theoretically justify the more elaborate structure needed for spontaneous CP violation or to construct U(1)-based flavor models in the spirit of Froggatt-Nielsen mechanisms.

\subsection{CP violation from spontaneously broken local U(1)}\label{sec:U1}
As is well known, a {\em local} U(1) broken by one scalar $S = f e^{i a/f}/\sqrt{2}$ leaves no physical particle $a$.
This is best seen by writing the action in terms of U(1)-neutral fields and performing the gauge transformation
$P \to e^{i q_P \alpha} P$, $S\to e^{i q_S \alpha} S$, $A_\mu\to A_\mu + \partial_\mu \alpha$ with $\alpha = -a/fq_S$:
this removes $a$ from couplings (so Yukawas become real), 
and gauge-invariant kinetic terms (being gauge-invariant) do not induce $\partial_\mu a$ couplings to particle currents.

\smallskip

A local U(1) is theoretically motivated and allows to write theories where CP is dynamically broken.
A minimal setup is a local U(1) spontaneously broken by two complex scalars 
$S_1 = f_1 e^{i a_1/f_1}/\sqrt{2}$ and $S_2 = f_2 e^{i a_2/f_2 }/\sqrt{2}$.
Then, Yukawa couplings can be given by the sum of at least two contributions, such as two powers of $S_1$ and $S_2$:
\beq Y(S_1,S_2) QUH = (c_1 S_1^{-q_{QUH}/q_{S_1}} + c_2 S_2^{-q_{QUH}/q_{S_2}} ) QU  H.\eeq
A gauge transformation $S_i \to e^{i q_{S_i} \alpha} S_i$  with $\alpha=-a/f$ shifts $a_1$ and $a_2$, allowing to remove
the combination $a = (q_{S_1} f_1 a_1 +q_{S_2} f_2 a_2)/f$ parallel to the shift,
leaving the orthogonal combination 
\beq \tau =  (q_{S_2} f_2 a_1  - q_{S_1} f_1 a_2)/f\qquad\hbox{where}\qquad f\equiv \sqrt{q_{S_1}^2 f_1^2 + q_{S_2}^2f_2 ^2}\eeq 
as a physical real scalar field.
The real scalar $\tau$ gets a mass around the symmetry breaking scale in the presence of potential terms that mix $S_1$ with $S_2$.
Furthermore, $\tau$ can have a mass mixing with the radial Higgs modes in $S_{1,2}$.
Its effective action is
  \beq \Lag_{\rm eff}= \left[y_1 \exp \left(-iq_{QUH} \frac{q_{S_2} f_2}{q_{S_1} f_1}\frac{\tau}{f} \right) + 
  y_2 \exp\left(iq_{QUH} \frac{q_{S_1} f_1}{q_{S_2} f_2} \frac{\tau}{f}\right)\right]QUH - \frac{m_\tau^2}{2}\tau^2+\cdots.
 %+ \sum_P  \left[ \frac{q_2 f_2}{q_1 f_1}-\frac{q_1 f_1}{q_2 f_2}\right]\frac{\partial_\mu\tau}{n}  q_P J^\mu_P
 \eeq 
This shows that  $\tau$ renders the Yukawa couplings complex with
only the CKM phase remaining as a physical CP-breaking parameter after $\tau$ settles to its minimum and is integrated out.
In the next section, we show that a qualitatively similar effective action arises from a more plausible  theory of spontaneous CP-breaking.

\subsection{CP violation from modular invariance}\label{sec:mod}
Modular invariance is the effective QFT description of a plausible super-string origin of CP,
see~\cite{2501.16427} for an intuitive summary.
For simplicity, we focus on the full modular group.
The theory below the string scale contains matter particles $P$ (the Higgs doublet $H$ and the fermions in the SM) 
alongside a special complex modulus scalar $\tau =\tau_R+i \tau_I $ 
such that the effective action is invariant 
under
\beq \tau\to \frac{a\tau+b}{c\tau+d},\qquad  P \to (c\tau+d)^{-k_P}P ,\qquad
Y(\tau)\to (c\tau+d)^{k_Y} Y(\tau)
\eeq
where $a,b,c,d$ are integers with $ad-bc=1$ forming the $\SL(2,\mathbb{Z})$ modular group.
The integers $k$ are called modular weights and
behave analogously to U(1) charges, as a modular transformation includes a phase rotation.
For instance, the $Y_{\rm u}(\tau) QUH$ interaction is modular invariant if $Y_{\rm u}(\tau)$ is a modular function
that transforms with modular weight $k_{Y_{\rm u}} = k_Q + k_U  + k_H$.

The modular-invariant kinetic terms have a non-canonical form
\beq \label{eq:LagKinMod}
\Lag_{\rm kin} =h^2 \frac{|\partial_\mu\tau|^2}{(-i\tau + i \bar\tau)^2} + \left[\frac{i}{2}\frac{\bar\psi \bar\sigma^\mu D_\mu\psi}{(-i\tau + i \bar\tau)^{k_\psi}} + \text{h.c.} \right]+
\frac{|D_\mu H|^2}{(-i\tau + i \bar\tau)^{k_H}} 
\eeq
that takes into account that modular transformations also operate rescalings.
The modulus $\tau$ is here dimensionless, but $h$ plays the same role as $f$ in section~\ref{sec:U1}.
Derivatives are replaced by modular-covariant derivatives that allow us to write modular-invariant kinetic terms by ensuring
 that $D_\mu P$ transforms like $P$ under modular transformations. The minimal form is\footnote{A non-minimal form, useful for dealing with anomalies,
is $D_\mu = \partial_\mu +i k \pi E_2(\tau) \partial_\mu \tau/6$~\cite{2406.02527},
% $1/\Im\tau$ is the difference between the two $E_2
that relies on the inhomogeneous anomalous modular transformation of the holomorphically regularised  $E_2$ Eisenstein function.}
\beq \label{eq:Dcovmin}
D_\mu = \partial_\mu +i k \frac{ \partial_\mu \tau}{-i\tau+ i \bar\tau}\,.\eeq
%  E_2(\tau) \to (c\tau+d)^2 E_2(\tau) - i\frac{6}{\pi} c(c\tau+d)\,.
Taking hermitian conjugates into account, each particle $P$ acquires a current coupling to $\partial_\mu \tau_R$ proportional to its modular weight $k$.
% tauR because kinetic terms in eq.\eq{LagKinMod} have the h.c.
 
For historical and pragmatic reasons, the literature focused on super-string compactifications that respect $N=1$ supersymmetry, 
possibly broken mildly below the string scale.
While supersymmetry is not relevant for our current purposes, it helps writing anomaly-free modular-invariant actions 
for chiral super-fields $\Phi$ that transform as $\Phi\to (c\tau+d)^{-k}\Phi $. 
The modulus $\tau$ too gets promoted to a chiral super-field.
The effective action for the theory is then described by the Kähler potential $K$ and by the super-potential $W$.
Their minimal form is
\beq \label{eq:SUSYaction}
K =- h^2 \ln(-i\tau+i\bar\tau)  +\sum_\Phi \frac{ \Phi^\dagger e^{2V} \Phi}{(-i\tau+i \bar\tau)^{k_\Phi}},
\qquad W = Y_{\rm u}(\tau) QU H_{\rm u}+\cdots.\eeq
%where $Y(\tau)$ is a modular function with weight $k_{QUH_u}$.
Here, $V$ stands for vector super-multiplets.
In the super-symmetric case, the K\"ahler geometry induces a non-trivial %holomorphic kinetic 
Kähler metric in field space,  
$K_i^{\bar \jmath} =  \partial^2 K/ \partial \Phi^i \partial \bar\Phi_{\bar \jmath}$,
such that the associated connections ensure modular invariance. 
 Translating from super-field  to component language,
the fermion derivative term covariant under field-space redefinitions is
$ D_\mu \psi^i = \partial_\mu \psi^i + \Gamma^i_{jk} (\partial_\mu \phi^j )\psi^k$.
The Christoffel symbols have non-vanishing entries for purely  holomorphic (or anti-holomorphic) indices, given by the
simpler expression  $\Gamma^i_{jk} = \left(K^{-1}\right)^i_{\bar m} \partial K^{\bar m}_{j}/\partial \phi^k$.
 The minimal Kähler kinetic term of eq.\eq{SUSYaction} gives a fermion kinetic term plus couplings of $\partial_\mu \tau_R$ to fermion currents
 in agreement with eq.\eq{Dcovmin}
 \beq \frac{1}{(2\tau_I)^{k}} \left(
\bar\psi  i \slashed{\partial} \psi - k \frac{ \partial_\mu\tau_R}{2\tau_I} \bar\psi \bar\sigma^\mu \psi\right).
 \label{eq:L-tau-fermions}
 \eeq
Namely, each fermion couples to the CP-breaking part of $\tau$ proportionally to its modular weight. 
The action has the general form of eq.\eq{Lagtau}, provided that
$\tau$ is reinterpreted as the real coordinate in field space along the cosmological evolution
of the multi-component complex $\tau$.

\section{Generation of the baryon asymmetry}\label{sec:cosmo}
We here show that, in theories of spontaneous CP-breaking  described by the effective action~\eq{Lagtau}, 
{\em the cosmological evolution of the scalar $\tau(t)$ --- as it possibly moves towards a CP-breaking minimum ---
can  generate the baryon asymmetry} of the universe.
The dynamics of a quantum field $\tau(t)$ is in general complicated, but a simple limit is enough for our purposes.
Assuming that the various interactions are fast enough to maintain kinetic thermal equilibrium during the $\tau(t)$ cosmological evolution,
the dynamics simplifies to compute chemical potentials $\mu_P$ for each matter particle $P$.
We recall that a chemical potential at temperature $T\gg\mu_P$ leads to a number asymmetry  
\beq\label{eq:Deltamu}
 \Delta_P =n_P -  n_{\bar P} = c_{\rm spin} d_P T^3  \frac{\mu_P}{T},\qquad c_{\rm spin}=
 \left\{ \begin{array}{ll}
1/6 & \hbox{fermion}\\
1/3 & \hbox{boson}
\end{array}\right.\eeq 
for a massless fermion or boson with $d_P$ degrees of freedom at temperature $T$.
So the baryon number density is 
\beq n_B = N_{\rm gen} (2\Delta_Q -\Delta_U -\Delta_D)/3 =   \mu_B T^2 /6\qquad\hbox{where} \qquad \mu_B\equiv 3(2\mu_Q-\mu_U-\mu_D).\eeq
We consider the action in eq.\eq{Lagtau}: chemical potentials are sourced by Yukawa couplings $Y(\tau)$, 
by  anomalous terms $\theta_{2,3}(\tau)$, and by current couplings $c_P(\tau)$.

\begin{enumerate}[a)]
\item Current terms
$ c_P (\partial_\mu \tau) J^\mu_P = \mu_P J_P^0$
contribute to the chemical potentials as 
\beq \mu_P = c_P \dot\tau,\eeq
% the sign is ok as \Lag = + \mu n 
given that $ J_P^0 $ is the number density of quanta of the field $P$~\cite{Cohen:1987vi}.

\item 
Yukawa couplings $Y(\tau)$, unlike current couplings, are not shift-invariant.
Their contribution to chemical potentials can be computed indirectly by noticing that physics is invariant under re-phasings such as eq.\eq{rephasing}.
Then the physical chemical potentials sourced by the quark Yukawa couplings $Y_{\rm u}$ and $Y_{\rm d}$
together with current terms
must be controlled by  the rephasing-invariant combination
\beq \label{eq:muYu}
\mu_{Y_{\rm u}} \equiv  (d\theta_{Y_{\rm u}}/d\tau + c_Q + c_U + c_H  )\dot\tau,\qquad
\mu_{Y_{\rm d}} \equiv  (d\theta_{Y_{\rm d}}/d\tau + c_Q + c_D - c_H  )\dot\tau\eeq
where we denoted the phases of the Yukawa couplings as $\theta_{Y_{\rm u}} = \Im\ln {Y_{\rm u}}$ and $\theta_{Y_{\rm d}} = \Im\ln {Y_{\rm d}}$.
Similar expressions hold for the lepton Yukawa couplings
\beq \label{eq:muYe}
\mu_{Y_{\rm e}} \equiv  (d\theta_{Y_{\rm e}}/d\tau + c_L + c_E - c_H  )\dot\tau,\qquad
\mu_{Y_{\nu}} \equiv  (d\theta_{Y_{\nu}}/d\tau + c_L + c_N +c_H  )\dot\tau\eeq
as well as for the right-handed neutrino mass term $M$:
we define $\mu_M = (d\theta_M/d\tau+2c_N)\dot\tau$ where $\theta_M$ is the phase of $M$.

\item The analogous re-phasing invariant combinations for anomalous terms, that lead to weak and strong sphaleron processes, are
\beq \label{eq:muS}
\mu_{S_2}= [-d\theta_2/d\tau + N_{\rm gen} (3 c_Q + c_L) ]\dot\tau,\qquad
\mu_{S_3} = [-d\theta_3/d\tau + N_{\rm gen} (2c_Q +c_U + c_D) ]\dot\tau.\eeq
\end{enumerate}
In the basis where $c_P=0$, the invariant parameters simply reduce to the time derivatives of the CP phases
\beq \mu_{Y_i} = \dot\theta_{Y_i},\qquad \mu_{S_i} = -\dot\theta_i.\eeq
In section~\ref{sec:evo}, we show how these parameters source chemical potentials
making explicit the connection between CP-violating phases and baryogenesis.

\begin{figure}[t]
$$\includegraphics[width=0.9\textwidth]{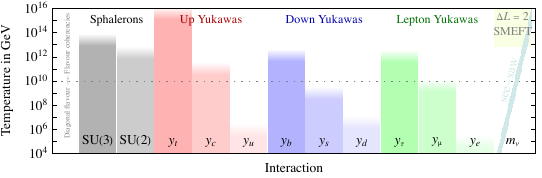}$$
\begin{center}
\caption{\em\label{fig:Tdec}
The shaded bands represent the range of temperatures where SM couplings equilibrate chemical potentials with rates fast enough to be in thermal equilibrium.
The last column considers $\Delta L=2$ interaction from atmospheric Majorana neutrino masses: 
the upper band is the misleading result of the $(LH)^2$ effective operator approximation; 
while the vertical bands represent the range assuming a right-handed neutrino, depending on its unknown mass.
Flavor effects undergo a qualitative change around the horizontal line:
at lower temperatures enough Yukawa rates are fast enough to suppress quantum coherence.
}
\end{center}
\end{figure}

\subsection{Evolution equations for chemical potentials}\label{sec:evo}
Eq.\eq{Deltamu} tells how small chemical potentials $\mu_P$ induce particle asymmetries $\Delta_P$. 
We now write the equations that describe how the various interactions induce a
time evolution $\dot \Delta_P = d\Delta_P/dt$.
We write the evolution equations for one generation only, 
without adding generation indices, 
because flavour violations are either fast enough to equilibrate the flavour asymmetries, or a more complicated density matrix formulation is needed
(see e.g.~\cite{hep-ph/9911315,1804.05066}).
We neglect usual thermal leptogenesis from CP violation in the decays of right-handed neutrinos $N$.
Including a Hubble rate $H$, the evolution equations are:
\begin{eqnsystem}{sys:dotDelta}
\dot\Delta_U + 3 H \Delta_U &=&  - S_{Y_{\rm u}}  - S_{S_3}\\
\dot\Delta_D + 3 H \Delta_D &=&  -S_{Y_{\rm d}} - S_{S_3}\\
\dot\Delta_Q + 3 H \Delta_Q &=&-S_{Y_{\rm u}} -S_{Y_{\rm d}} - 3 S_{S_2} - 2 S_{S_3}\\
\dot\Delta_L + 3 H \Delta_L &=& - S_{Y_{\rm e}}- S_{S_2} - 2S_{\Delta L=2}\\
\dot\Delta_E + 3 H \Delta_E &=&- S_{Y_{\rm e}} \label{eq:DeltaE}\\
\dot\Delta_H + 3 H \Delta_H &=&  N_{\rm gen}(-S_{Y_{\rm u}} +S_{Y_{\rm d}}+ S_{Y_{\rm e}} -2  S_{\Delta L=2}).
\end{eqnsystem}
%Chemical potentials for vectors are loop suppressed.
Each interaction $I$ contributes to a term $S_I$. 
All the $S_I$ have a common structure, being given by the space-time 
density $\gamma_I$
of decays or scatterings induced by $I$, times the sum of the chemical potentials
of the particles involved in the process, including a source term $\mu_I \propto \dot\tau$ from each interaction.
These are the factors computed in eq.\eq{muYu},\eq{muYe},\eq{muS}. Explicitly:
\begin{itemize}
\item The terms induced by the Yukawa couplings are
\beq S_{Y_{\rm u}}= \gamma_{Y_{\rm u}} \frac{\mu_Q+\mu_U+\mu_H - \mu_{Y_{\rm u}} }{T},\qquad
S_{Y_{\rm d}}= \gamma_{Y_{\rm d}} \frac{\mu_Q+\mu_D-\mu_H - \mu_{Y_{\rm d}} }{T},
 \eeq
\beq \label{eq:SYnue}
S_{Y_{\rm e}}=\gamma_{Y_{\rm e}} \frac{\mu_L+\mu_E-\mu_H - \mu_{Y_{\rm e}} }{T},\qquad
S_{Y_{\nu}}= \gamma_{Y_{\nu}} \frac{\mu_L+\mu_N+\mu_H - \mu_{Y_{\nu}} }{T}. \eeq
If all particles are massless, the rates are $\gamma_{Y_{\rm u}}\approx \alpha_3 |Y_{\rm u}|^2 T^4$,
$\gamma_{Y_{\rm e}}\approx \alpha_2 |Y_{\rm e}|^2 T^4$, etc., corresponding to $2\to 2$ scatterings.
The various SM interactions reach thermal equilibrium in the temperature range plotted in fig.\fig{Tdec}.

\item The terms induced by weak $\SU(2)_L$ and strong $\SU(3)_c$ sphalerons are
\beq  S_{S_2}=\gamma_{S_2} \frac{N_{\rm gen}(\mu_{L} + 3\mu_{Q}) - \mu_{S_2}}{T},\qquad
S_{S_3}= \gamma_{S_3} \frac{N_{\rm gen}(2\mu_Q+\mu_U+\mu_D )- \mu_{S_3} }{T}
\eeq
with rates $\gamma_{S_3} \approx 100\alpha_3^5 T^4$ and $\gamma_{S_2} \approx 10\alpha_2^5 T^4$.

\item The term $S_{\Delta L=2}$ denotes interactions that violate lepton number by 2 units,
arising from virtual exchange of right-handed neutrinos.
Two left-handed leptons   and two Higgs are involved, leading to
\beq \label{eq:SDeltaL2}
S_{\Delta L=2} = \gamma_{\Delta L=2} \frac{2(\mu_L+\mu_H) - \mu_{LLHH}}{T}\qquad\hbox{where}\qquad
\mu_{LLHH} = 2 \mu_{Y_\nu} - \mu_M.\eeq
\end{itemize}

\subsection{$\Delta L=2$ interactions}
In this section, we justify the form of the $S_{\Delta L=2}$ term and discuss its  $\gamma_{\Delta L=2}$ rate.
We assume neutrino masses $m_\nu$ of Majorana type, that violate lepton number.

\smallskip

At temperatures much below the mass $M$ of right-handed neutrinos, 
their effects get approximated as the effective neutrino Majorana mass operator 
$(LH)^2/2\Lambda$ suppressed by a scale
$\Lambda = v^2/m_\nu \lesssim 6~10^{14}$ where $v=174\GeV$.
In the operator limit, $S_{\Delta L=2}$ is immediately computed, and
the interaction rate $\gamma_{\Delta L=2}$ is 
faster than the Hubble rate at $ T \gtrsim 4\pi \sqrt{d_{\rm SM}}\Lambda^2/\bp \approx 6~10^{12}\GeV$ (see e.g.~\cite{2006.03148}).
%This would provide the needed $B-L$ violation with the needed decoupling temperature.

However, the effective operator approximation can be misleading.
So we assumed the most plausible full theory, where the neutrino mass operators are mediated as
$1/\Lambda = Y_\nu^2/M$ by tree-level exchange of right-handed neutrinos $N$,
with Yukawa couplings $Y_\nu (\tau)$ and mass terms $M (\tau)$
as anticipated in eq.\eq{Lagtau}.
The $\Delta L=2$ interaction term needs to be decomposed as
\beq 2S_{\Delta L=2}=S_{Y_{\nu}}+2S_{\Delta L=2}^{\rm off-shell}\eeq
where $S_{Y_\nu}$ (defined in eq.\eq{SYnue} but not yet used) 
has rate $\gamma_{Y_\nu}=\gamma_D/2$
dominated by on-shell $N \leftrightarrow LH, \bar L H^*$ decays of the massive $N$ with width $\Gamma_N$.
The space-time density of $N$ decays is:
\beq \gamma_D= n_N^{\rm eq} \med{\Gamma_N} =n_N^{\rm eq} \frac{{\rm K}_1(M/T)}{{\rm K}_2(M/T)}\Gamma_N,\qquad
\Gamma_N = \frac{|Y_\nu|^2 M}{8\pi} \hbox{~at rest}.
\eeq
 We can here neglect $2\to 2$ scatterings, as massive $N$ can decay.
To avoid overcounting, the remaining term $S_{\Delta L=2}^{\rm off-shell}$ 
only includes the contribution due to off-shell scatterings
and is thereby sub-leading for perturbatively small values of $Y_\nu$.
It becomes relevant at  $T \ll M$, when the number density $n_N^{\rm eq}$ of right-handed neutrino gets Boltzmann suppressed.
See~\cite{hep-ph/0310123} for a discussion of these issues in the context of leptogenesis.

\smallskip

Next, the evolution eq.s~(\ref{sys:dotDelta}) need to be complemented adding an extra equation for $N$,
\beq \label{eq:DeltaN}\dot\Delta_N + 3 H \Delta_N =- S_{Y_{\nu}} - 2S_M,\qquad
S_M = \gamma_M \frac{2\mu_N - \mu_M}{T}.\eeq
In eq.\eq{DeltaN} we treated the right-handed neutrino mass as
a very fast interaction term $\gamma_M \sim M T^3$, leading to a nearly-equilibrium solution
$S_M=0$ and thereby to $\mu_N = \mu_M/2$.
Inserting this into the other eq.s~(\ref{sys:dotDelta}) leads to eq.\eq{SDeltaL2}.

\medskip

Having clarified the theory, we next discuss the numerical value of $\gamma_{\Delta L=2}\simeq\gamma_D/4$.
Neutrino oscillation experiments observed two different neutrino mass splittings~\cite{2503.07752}
\beq |\Delta m^2_{32}| = (2.50\pm0.02)10^{-3} \eV^2 ,\qquad \Delta m^2_{21}= (7.4\pm 0.2)10^{-5}\eV^2 .
\eeq
%Ignoring the possibility of quasi-degenerate neutrinos, 
We focus on the plausible minimal `normal ordering' possibility for neutrino masses
\beq 
m_{\nu_1} \ll m_{\nu_2}  = m_{\rm sun}  \simeq r \sqrt{\Delta m^2_{21}}\ \ll m_{\nu_3} = m_{\rm atm}  \simeq  r \sqrt{ |\Delta m^2_{32}| }\eeq
having included a renormalisation factor $r$, equal to $r\approx 1.2$ at $10^{10}\GeV$ in the SM (see~\cite{hep-ph/0606054} for a review).
Furthermore, we focus on the  minimal possibility for right-handed neutrinos:
one right-handed neutrino mediates $\tilde{m}=m_{\rm atm}$, another one mediates $\tilde{m}=m_{\rm sun}$.
The decay rate at $T=M$ is
\beq \frac{\Gamma_N}{H} \approx \frac{\tilde{m}}{m_*},\qquad
m_* = \frac{256\sqrt{d_{\rm SM}} v^2}{3 M_{\rm Pl}} \approx 2.3\,{\rm meV}.
\eeq
This means that:
\begin{itemize}
\item[atm)] The decay rate of the right-handed neutrino that mediates the larger $m_{\rm atm} \gtrsim m_*$ 
is inevitably mildly in thermal equilibrium within a temperature range from a fraction of its mass to 10 times its mass.
The unknown mass $M$ is constrained to be below $10^{15}\GeV$ by perturbativity of Yukawa couplings $Y_\nu$.
Fig.\fig{Tdec} illustrates the situation.

\item[sun)] The decay rate of the right-handed neutrino that mediates the smaller $m_{\rm sun} \lesssim m_*$ is mildly below thermal equilibrium
at any temperature, and reaches the maximal $\Gamma_N/H$ at $T \sim M$.
\end{itemize}
A different plausible possibility for $B-L$ violation is that $\U(1)_{B-L}$ is a gauge symmetry with gauge coupling $g_{B-L}$,
spontaneously broken at some large scale leading to a vector boson with mass $M_{B-L}$.
This situation is for example predicted by SO(10) gauge unification, while renormalizable minimal SU(5) interactions accidentally conserve $B-L$.
This $B-L$ violation  decouples 
at $T_{\rm dec} \sim d^{1/6}_{\rm SM} (M_{B-L}/g_{B-L})^{4/3} /\bp^{1/3}$.
In conclusion, $B-L$ can be violated by neutrino masses or by some other new physics.

\subsection{Equilibrium solutions}\label{sec:eqsol}
To see how the CP-violating $\tau$ dependence of Yukawa couplings and sphalerons
can source baryogenesis,
it is useful to start considering equilibrium solutions to eq.s~(\ref{sys:dotDelta}).
The full solutions will later be approximated as equilibrium solutions at some decoupling moment.

Equilibrium solutions amount to set $\dot\Delta_P=0$ and $H =0$.
Then eq.s~(\ref{sys:dotDelta}) become a system of linear equations for the $\mu_P$, with coefficients that depend on the rates $\gamma_I$.
Such equations are almost equivalent to simply demanding that all interaction terms $S_I$ vanish  separately.
If fully correct, this would lead to the following system of equations for the chemical potentials, with coefficients that do not depend on rates $\gamma_I$:
\beq\label{eq:equil}
\left\{\begin{array}{ll}
\hbox{$\Delta L =2$ violation}: &\mu_{LLHH}= 2\mu_L + 2\mu_H\\
\hbox{Lepton Yukawa $EL\bar{H}$}: & \mu_{Y_{\rm e}}=\mu_E+\mu_L-\mu_H\\
\hbox{Down-quark Yukawa $DQ\bar{H}$}: &\mu_{Y_{\rm d}}=\mu_D+\mu_Q-\mu_H\\
\hbox{Up-quark Yukawa $UQ{H}$}: &\mu_{Y_{\rm u}}=\mu_U+\mu_Q+\mu_H\\
\hbox{$\SU(3)_c$ sphalerons}:& \mu_{S_3} = N_{\rm gen} (2\mu_Q +\mu_U + \mu_D)\\
\hbox{$\SU(2)_L$ sphalerons}:&\mu_{S_2}=N_{\rm gen}(3\mu_Q+\mu_L)\\
\hbox{$\U(1)_Y$ conservation}: &0=N_{\rm gen}(\mu_Q-2\mu_U+\mu_D-\mu_L+\mu_E)+2\mu_H .
\end{array}\right.\eeq
However, these simpler equations are incompatible whenever $\bar\theta_3(\tau)$ depends on $\tau$
because this leads to $\mu_{Y_{\rm u}} + \mu_{Y_{\rm d}}\neq  \mu_{S_3}$, implying
that the true equilibrium solution involves a dynamical balance where strong sphalerons create an asymmetry that 
gets cancelled by $Y_{\rm u}$ and $Y_{\rm d}$ interactions.\footnote{A modular symmetry can solve
the strong CP problem if the modular weights are chosen such that the rephasing-invariant combination
$\bar\theta_3=\theta_3(\tau)+\arg\det Y_{\rm u}(\tau)Y_{\rm d}(\tau)$ vanishes~\cite{2305.08908,2404.08032,2406.01689}.
In the present context, this condition implies $\mu_{Y_{\rm u}} + \mu_{Y_{\rm d}}=\mu_{S_3}$, such that
modular baryogenesis can be generated successfully while avoiding the issue discussed here.}
This is why the equilibrium solution then depends on the rates $\gamma_I$, not just on the $\mu_I$ sources.
In practice, in the SM the bottom Yukawa mediates a significantly slower interaction than
the top Yukawa and than the strong sphalerons, $\gamma_{Y_{\rm d}} \ll \gamma_{S_3} , \gamma_{Y_{\rm u}}$.
As a result, a good approximation is obtained by simply omitting the down-quark Yukawa equation from eq.s~(\ref{eq:equil}).
This gives the equilibrium solution
\beq \mu_{B-L}^{\rm eq} \simeq \frac{72 \mu_{Y_{\rm u}}+9 \mu_{Y_{\rm e}}}{11} - \frac{79}{22}\mu_{LLHH}+\frac{28}{33} \mu_{S_2}-\frac{19}{11}\mu_{S_3}.
\eeq
This shows that a baryon asymmetry is sourced by a
time dependence of the phase of Yukawa couplings (even of the lepton Yukawas only) or of the sphalerons.

\subsection{Solving the evolution equations}\label{sec:solving}
We assume that the total energy density $\rho_{\rm SM}$ of relativistic SM (and beyond) particles arises from an inflaton $\phi$ (possibly identified as $\tau$)
with a decay rate $\Gamma_\phi$. The process is described by the usual reheating equations:
\beq \label{eq:cosmoevo}
\frac{d\rho_\phi}{dt}+3H \rho_\phi= -  \Gamma_{\phi} \rho_\phi,\qquad
\frac{d\rho_{\rm SM}}{dt}+4H\rho_{\rm SM} =\Gamma_{\phi} \rho_\phi
\eeq
where the Hubble rate $H$ at temperature $T$ and time $t$ is
\begin{equation}\label{eq:Hubble}
H^2=\frac{\rho_{\rm SM}+\rho_{\phi}}{3 \bp^2},\qquad   \rho_{\rm SM} = \frac{\pi^2}{30}d_{\rm SM}(T)T^4.
\end{equation}
The number of degrees of freedom in the SM much above the weak scale is $d_{\rm SM}\approx 106.75$.
The evolution equations can be rewritten by switching from time $t$ to scale factor $a$ to temperature $T$ using
\beq \frac{d}{dt} = H \frac{d}{d\ln a} = - ZH \frac{d}{d\ln T} ,\qquad
Z= 1 - \frac14 \frac{\Gamma_{\rm SM}\rho_\phi}{H\rho_{\rm SM}} \eeq
where $Z\neq 1$ describes entropy injection from inflaton decays.
The equation for $\rho_{\rm SM}$ can be dropped (see e.g.~\cite{hep-ph/0310123})
and time derivatives simplify as
\beq \dot\Delta_P + 3 H \Delta_P = -c_{\rm spin} d_P T^3 H \left[Z \frac{d}{d\ln T}\frac{\mu_P}{T} +3 (Z-1) \frac{\mu_P}{T}\right]\eeq
and the equations for $\Delta_P$ can be rewritten as equations for $\mu_P/T$.
For example, the equation\eq{DeltaE} for $\Delta_E$ becomes
\beq \frac{d_E}{6}\left[Z \frac{d}{d\ln T}\frac{\mu_E}{T} +3 (Z-1) \frac{\mu_E}{T}\right]=
\frac{\gamma_{Y_{\rm e}}}{HT^3} \frac{\mu_L+\mu_E-\mu_H - \mu_{Y_{\rm e}} }{T}.\eeq
An interaction $I$ is in thermal equilibrium when the dimensionless ratio $\gamma_I/ T^3 H$ is much larger than unity,
implying the smallness of the combination of chemical potentials it multiplies in the evolution equations.
Assuming that only $\gamma_{\Delta L=2}$ can be small, while all other $\gamma_I$ are in equilibrium,
the system of equations~(\ref{sys:dotDelta}) can be approximated
as one equation for $B-L$:
\beq \label{eq:evoB-L}
Z  \frac{d}{d\ln T}\frac{\mu_{B-L}}{T}+3(Z-1)  \frac{\mu_{B-L}}{T}=   -  \frac{66}{237}\frac{\gamma_{\Delta L=2}}{HT^3} \frac{\mu_{B-L} - \mu_{B-L}^{\rm eq} }{T} .
\eeq
This shows that $\mu_{B-L}$ freezes at a value $\mu_{B-L}^{\rm dec} \sim \mu_{B-L}^{\rm eq}(T_{\rm dec})$,
where $T_{\rm dec}$ is the decoupling temperature of $\Delta L=2$ interactions
(unless the inflaton is still releasing entropy, $Z\neq 1$).
Next, SM equilibration conditions imply 
$\mu_B = \frac{28}{79}\mu_{B-L}^{\rm dec}$
such that 
\beq\label{eq:nBs}
 \frac{n_B}{s} = \frac{15}{4\pi^2 d_{\rm SM}} \frac{\mu_B}{T}.
\eeq
This ratio remains constant and the current entropy density is $s \approx 7.04 n_\gamma$. 
Omitting model-dependent order unity factors one has $\mu_B \sim \dot\tau$ at decoupling.
One then needs to solve the field equation for $\tau$, that can be approximated as
\beq \ddot\tau+3H \dot\tau = - m_\tau^2( \tau - \tau_f) \eeq
where the potential has been approximated  as quadratic around the minimum at $\tau_f$.\footnote{We do not attempt a full model-dependent analysis for the following reasons.
In the modular context, super-gravity potentials with a CP-breaking  local minimum for $\tau$ have been written in~\cite{2201.02020}.
However this minimum has a Planckian negative vacuum energy.
Finding a Minkowski vacuum for moduli is a notoriously complicated issue in string compactifications.
Assuming it can be solved, some authors proposed identifying
the modulus $\tau$ as the inflaton, either by tuning the top of the potential such that a second derivative $V''$ is small~\cite{2411.18603} 
(see also~\cite{2405.06497,2405.08924}), or
by conjecturing a Starobinski-like potential, maybe derivable from $D$-terms~\cite{2407.12081}.}  
The mass term $m_\tau$ can include thermal contributions coming from the interactions of $\tau$ with the SM bath,
a back-reaction from the small asymmetries,
and curvature contributions from a non-minimal couplings of $\tau$ to gravity.
The solution to the $\tau$ field equation yields a small $\dot\tau/\tau \simeq - m_\tau^2/3H$ in the Hubble-damped regime $H(T)\gg m_\tau$.
At lower temperatures $T_{\rm osc} $ one has $H(T_{\rm osc} )\sim m_\tau$ and the scalar $\tau$ begins oscillating around its minimum while red-shifting as matter.
So a maximal $\dot \tau \sim m_\tau$ arises, leading to
\beq\label{eq:YBnaive}
 \frac{n_B}{n_\gamma} \lesssim \frac{\mu_{B-L}^{\rm dec}}{T_{\rm dec}}  \lesssim \frac{T_{\rm dec}}{\bp} .\eeq
The observed baryon asymmetry is reproduced if $T_{\rm dec}\gtrsim 10^{10}\GeV$, similarly to~\cite{Cohen:1987vi}.
Assuming radiation domination, the full solution for $\tau$ as function of temperature $T$,
given an initial value $\tau_i$ and an initially negligible $\dot \tau_i$ at an initial temperature $ T_i$,  is~\cite{Cohen:1987vi}
\beq \label{eq:tauevo}
\tau(T) =\tau_f + (\tau_i-\tau_f) \sqrt{\frac{T}{T_i}}  \frac{ J_{1/4}(m_\tau/2H(T)) }{J_{1/4}(m_\tau/2H(T_i)) }\eeq
where $J_{1/4}$ is a Bessel function.

We assume that the scalar $\tau$, after having sourced the baryon asymmetry, finally decays leaving no other effects.
The viability of this assumption will be verified in section~\ref{sec:decay}.

\subsection{Modular baryogenesis at $T \sim 10^{11}\GeV$}\label{sec:T11}
The naive estimate of eq.\eq{YBnaive} for the amount of baryon asymmetry 
suggests that some $B$ or $L$-violating interaction must decouple around $T_{\rm dec}\sim10^{11}\GeV$.
A plausible possibility is that $\Delta L =2$ interactions decouple at $T_{\rm dec}$ given by the mass $M$ of the lighter right-handed neutrino that mediates either the `solar' or the `atmospheric'
neutrino mass splitting. %  as  $m_\nu = y_1^2 v^2/M_1$.
The modulus $\tau$ can have a time dependence at $T_{\rm dec}$  if it is
 light enough, $m_\tau\sim H \sim T_{\rm dec}^2/\bp \sim 10 \TeV$.
 
 \begin{figure}[t]
$$\includegraphics[width=\textwidth]{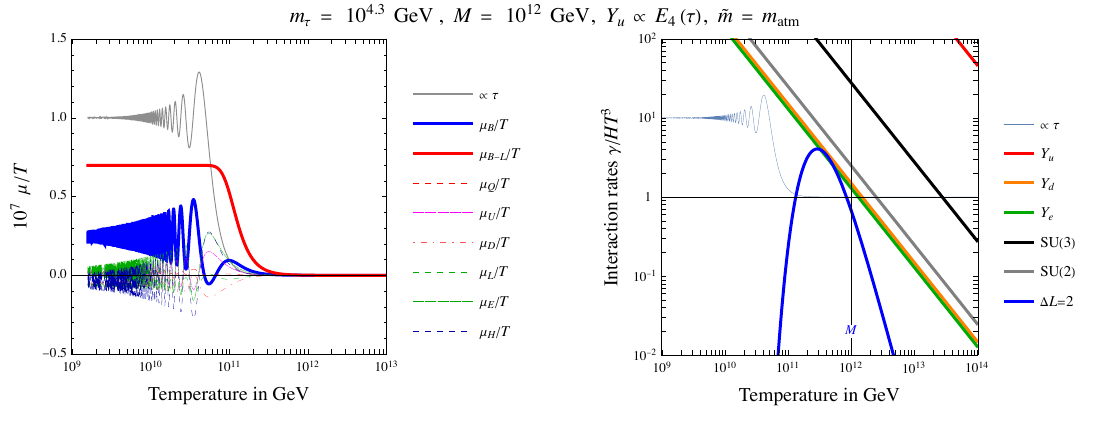}$$
\begin{center}
\caption{\em\label{fig:evo} 
Sample evolution of the chemical potentials $\mu_P/T$ (left) and of rates $\gamma_I/H T^3$ (right).
The CP-breaking scalar $\tau$ is relatively light, and both panels also include its evolution, in arbitrary units.
Baryogenesis is sourced around $10^{11}\GeV$
by the varying phase of up Yukawa coupling, assumed to be proportional to $E_4(\tau)$.
At this temperature all rates are in thermal equilibrium, except for the $\Delta L=2$ rate,
so that eq.\eq{evoB-L} for $\mu_{B-L}$ would be enough.
}
\end{center}
\end{figure}

Fig.\fig{evo} shows a numerical example.
We assume $M =10^{12}\GeV$, $m_\tau \sim 10\TeV$, $\tilde{m} = m_{\rm atm}$.
Given that we ignore flavour issues, we see no point in writing down a one-flavour toy model:
we instead assume that the only CP violation is a up quark Yukawa coupling $Y_{\rm u}$ proportional to the modular form with lowest weight 4, 
the modular Eisenstein form $E_4(\tau)$.
All other parameters are assumed to be $\tau$-independent.\footnote{Eq.\eq{L-tau-fermions}
implies extra current contributions to $\mu_{Y_{\rm u}}$ proportional to $k_{QUH}=4$, that we ignore.
A $\tau$ dependence of other Yukawa couplings or of $\theta_{2,3}$ would give similar results, 
as all these interactions are in thermal equilibrium when $\tau$ starts evolving.
Concerning the theta terms, a toy one-generation minimal setup where the effective field theory content has no modular/gauge/gauge anomalies
implies that $Y_{\rm u}$ and $Y_{\rm e}$ have opposite modular weights, a relation that can be avoided by allowing for
$\SL(2,\mathbb{Z}) \U(1)_Y^2$ anomalies, compensated by heavy fields.}

We assume that $\tau$ moves along constant $\Im\tau = 1$
from a CP-conserving point $\Re\tau_i = 0$ 
at an initial temperature $T_i \gg T_{\rm dec}$ to a CP-violating point $\Re\tau_f = 1/4$ as in eq.\eq{tauevo},
such that the phase of $Y_{\rm u}$ changes by order unity.
A more generic motion would similarly generate baryogenesis, including the opposite motion
from a CP-violating point to a CP-conserving point.
So other moduli that end up not breaking CP can generate baryogenesis too.\footnote{We neglect the
 time dependence of the absolute value of the Yukawa couplings (possibly but not necessarily related to a time-dependence 
of the `dilaton-like' component $\sim\tau_I$ of the complex scalar $\tau$).
Formally, this time dependence contributes as an imaginary part to the chemical potentials.
Physically, this means a symmetric contribution to particle and anti-particle creation due to the rolling of the scalar field.
Ordinary tree-level $\tau$ decays similarly contribute to the
transfer rate of energy from the scalar field to the SM bath.
We also neglect possible CP-violating decays of $\tau$ quanta.
These could arise when $\tau$ has at least two different decay modes.
CP-violation in $\tau/\tau^*$ mixing is expected to be negligible because 
the two eigenstates of the complex $\tau$ field are expected to have significantly different masses,
as they describe different aspects of the compactification geometry.
}
We fix $h =\bp$ as suggested by string theory, but any large value would give the same final amount of baryogenesis.
We assume that SM particles dominate the cosmological energy density.
The left panel of fig.\fig{evo} shows that the evolution of the chemical potentials $\mu/T$ (time flows from right to left)
is well approximated by the simple effective equation of eq.\eq{evoB-L}.
We neglected entropy release from the inflaton, setting $Z=1$.

A different related possibility is that the reheating temperature after inflation is $T_{\rm RH}\sim T_{\rm dec}$:
in this case $Z<1$ and $\tau$ can be time-dependent at this stage either because it is the inflaton
or, more in general, because the oscillating inflaton $\phi(t)$ induces
some time dependence in other scalars such as $\tau(t)$, or $\phi$ contains a component in field space along $\tau$.
Even in this case, $\tau$ must be relatively light.

%%\nnn{The scalar $\tau$ is generically unstable, with a decay width $\Gamma_\tau \sim m_\tau^3/h^2$, where $h$ is its decay constant.
%%So far its value was essentially irrelevant.
%%Having here assumed a light $\tau$, a Planckian value $h\sim \bp$ implies a slow decay that risks re-heating the universe
%%and thereby suppressing the baryon asymmetry.}

\begin{figure}[t]
$$\includegraphics[width=\textwidth]{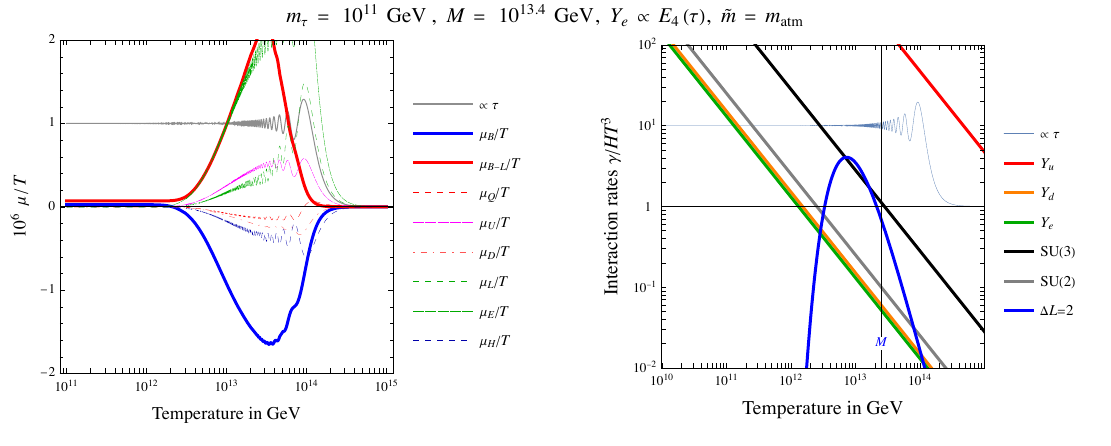}$$
\begin{center}
\caption{\em\label{fig:evoatm} 
Sample evolution of the chemical potentials $\mu_P/T$ (left) and of rates $\gamma_I/H T^3$ (right).
The CP-breaking scalar is now heavier, $m_\tau=10^{11}\GeV$.
Baryogenesis is sourced around $10^{14}\GeV$
by the varying phase of the charged lepton Yukawa coupling, assumed to be proportional to $E_4(\tau)$.
Some rates are not in thermal equilibrium,
}
\end{center}
\end{figure}

\begin{figure}[t]
$$\includegraphics[width=\textwidth]{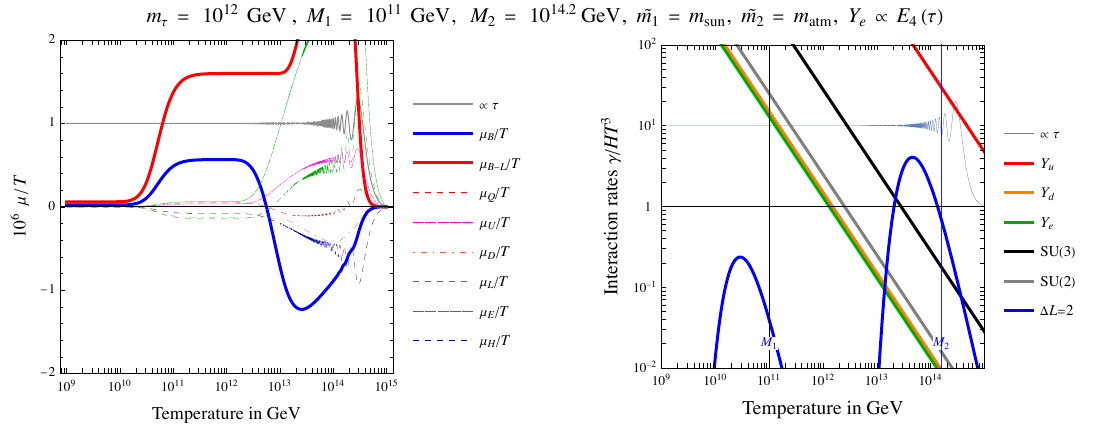}$$
\begin{center}
\caption{\em\label{fig:evoSun} 
Sample evolution of the chemical potentials $\mu_P/T$ (left) and of rates $\gamma_I/H T^3$ (right).
In this example the too large baryon asymmetry generated around the mass of the heavier right-handed
neutrino is later partially washed out by $\Delta L=2$ interactions mediated by the lighter right-handed neutrino.
}
\end{center}
\end{figure}

\subsection{Modular baryogenesis at $T \gg 10^{11}\GeV$}\label{sec:T>>11}
A heavier $\tau$  is time-dependent at a higher temperature.
The maximal temperature during the big bang is the reheating temperature, constrained to be $T_{\rm RH}\lesssim 0.003\bp$
by the non-observation of inflationary tensor modes~\cite{Planck:2018vyg}.
The modulus $\tau$ can have a time dependence provided it is not much heavier than the inflaton\footnote{A modulus $\tau$
with mass comparable to the inflationary Hubble scale can also lead to cosmo-collider signatures related to the chemical potentials (see e.g.~\cite{1805.02656}).}
\beq m_\tau \lesssim T_{\rm RH}^2/\bp\sim 10^{13}\GeV.\eeq
The amount of modular baryogenesis generated around the maximal $T_{\rm RH}$
is naively too large, $n_{B}/n_\gamma \sim T_{\rm RH}/\bp\sim 10^{-3}$.
It can be suppressed down to the observed value by assuming that, while $\dot\tau$ is maximal,
the inflaton is still releasing entropy  (described by the factor $Z-1$) and/or
fast $\Delta L=2$ interactions partially wash-out the baryon asymmetry.

Fig.\fig{evoatm} shows a numerical example that leads to the correct $n_B/n_\gamma$. 
We assume $m_\tau=10^{12}$~GeV, the maximal value consistent with iso-curvature bounds discussed later in eq.\eq{isoc}.
The temperatures involved are so high that the $Y_{\rm e,d}$ interactions and the sphalerons are mildly out of equilibrium.
This prevents approximating the baryogenesis dynamics as a single equation for $\mu_{B-L}$.
We solve the complete system of evolution equations maintaining the same initial and final $\tau$ as in the previous section.
Here, we assume that the only source of CP violation is a $\tau$-dependent charged lepton Yukawa $Y_{\rm e}$ proportional to the modular form $E_4(\tau)$.
This serves to illustrate that CP-violating lepton phases alone can generate the baryon asymmetry.
The key point is assuming a right-handed neutrino that mediates $\tilde{m}=m_{\rm atm}$
with mass $M\sim 10^{13}\GeV$, slightly below the temperature at which $\dot\tau$ reaches its maximum.
Consequently, the $\Delta L=2$ interactions significantly suppress the initially excessive baryon asymmetry 
bringing it down to the observed value.

The $\Delta L=2$ washout can be studied more in general.
Neglecting flavour issues and CP-violation in $N_1$ decays (that can generate baryogenesis via thermal leptogenesis if $M_1 \gtrsim 10^9\GeV$),
the lighter right-handed neutrino $N_1$ with mass $M_1$ suppresses a pre-existing  $B-L$ asymmetry generated at $T \gg M_1$
by a multiplicative factor $e^{-2.5 \tilde{m}_1/m_*} $.
If $\tilde{m}_1 =m_{\rm sun}$ this washout factor equals $ 10^{-5.3}$, irrespectively of $M_1$, providing the desired level of suppression.
Fig.\fig{evoSun} provides a numerical example that realises this possibility.
If instead $\tilde{m}_1 =  m_{\rm atm}$, $N_1$ interactions 
suppress the $B-L$ asymmetry by $\sim 10^{-30}$, down to a negligible value.\footnote{We used the rates computed in~\cite{hep-ph/0310123}.
The dominant suppression arises around $T \sim (0.1- 1)M$. Thermal corrections are relevant at larger $T$.
Loop corrections are a few $\%$~\cite{1106.2814,1112.1205}.
%At $T\gg M$ $\gamma_D$ gets suppressed by on-shell. 
%The properly subtracted off-shell rate is $\Gamma^{\rm sub} \sim y^4 T$ is negligible~\cite{hep-ph/0310123}.
%The gauge scattering is smaller than $\gamma_D$.
}
Taking flavour into account can significantly reduce the washout, but makes it highly model-dependent, as first studied in~\cite{hep-ph/9911315}.
We must now consider the three $B/3-L_{e,\mu,\tau}$ asymmetries, described by a $3\times 3$ density matrix. 
The modulus $\dot\tau$ typically sources all three asymmetries, in comparable but different amounts.
A right-handed neutrino heavier than $\approx 10^{11}\GeV$ 
acts as a polariser, washing the asymmetry only in the direction in flavour space parallel to its Yukawa couplings,
because it acts when the interactions due to charged lepton Yukawa couplings $y_{e,\mu,\tau}$ are out of thermal equilibrium.
A right-handed neutrino lighter than $\approx 10^9\GeV$ separately suppresses each $B/3 - L_{e,\mu,\tau}$,
because fast $y_{\mu,\tau}$ interactions remove the flavour coherences in the density matrix.
An intermediate situation arises in the intermediate mass range where only $y_\tau$ is in thermal equilibrium.

The right-handed neutrino $N_1$ can carry a lepton number asymmetry when it is very light, $M_1 \ll T$, such that $M_1$ operates a slow wash-out.
This effect is accounted by eq.\eq{DeltaN}. Nevertheless, we verified that the final result is well approximated by eq.\eq{SDeltaL2}
given that  the Yukawa coupling $Y_{\nu 1}$ is also suppressed and the asymmetry remains small.
%\MD{In this scenario, one could insist on describing the asymmetry of the lighter neutrino with eq. \ref{eq:DeltaN} at high temperatures $T\sim M_2>>M_1$ when $M_1$ is negligibile and the condition to replace it with eq. \ref{eq:SDeltaL2} is not satisfied. However, the respective Yukawa $Y_{\nu 1}$ is also suppressed for $T\gg M_1$ and the asymmetry remains tiny. We have checked numerically that the use of eq. \ref{eq:SDeltaL2} is valid also in this case.} 

Finally, as discussed in the next section, two extra effects can contribute to the desired partial suppressing of the baryon asymmetry:
i)~slow $\tau$ decays that reheat the universe;
ii)~heavy $m_\tau \sim H_{\rm infl}$ with suppressed motion after inflation.

\section{$\tau$ decays and iso-curvature perturbations}\label{sec:decay}
We assumed that the scalar $\tau$, after having sourced the baryon asymmetry at $T_{\rm osc} \sim T_{\rm dec}$, later decays without leaving extra effects.
The scalar must decay when the universe temperature is $T_{\rm decay} \lesssim T_{\rm osc}$,
before that its cosmological abundance $\Omega_\tau (T) $   becomes of order unity.
This ensures that the universe is not reheated in a way that would 
dilute the baryon asymmetry, although a partial wash-out is  allowed.
In the minimal model $\tau$  decays into SM particles 
with width $\Gamma_\tau \sim m_\tau^3/h^2$, where $h$ is the $\tau$ decay constant. 
So the $\tau$ decay happens at $T_{\rm decay}/T_{\rm osc} \sim m_\tau/h$.
Its energy fraction $\Omega_\tau (T_{\rm osc})\sim h^2/\bp^2$ grows up to $\Omega_\tau (T_{\rm decay})\sim h^3 /\bp^2m_\tau$,
implying that  a decay fast enough to make reheating negligible needs 
\beq \label{eq:fastdecay} h\lesssim \bp^{2/3}m_\tau^{1/3}.\eeq
The precise value of $h$ negligibly affects the baryon asymmetry.

\medskip

%\MD{Such scenario naturally arises for $m_\tau < h \ll \bp$, when the energy fraction of $\tau$ is suppressed during oscillations ($H\sim m_\tau$, $T_{\rm osc}\sim\sqrt{m_\tau \bp}$) by $h^2/6\bp^2$ and remains small at decay ($H\sim \Gamma_\tau$, $T_\Gamma\sim m_\tau T_{\rm osc}/h $) if also $h\ll \bp^{2/3}m_\tau^{1/3}$.}}
%
The baryon asymmetry is approximately proportional to $\dot\tau$ at the decoupling temperature $T_{\rm dec}$. 
Consequently, {\em iso-curvature perturbations} are generated if, during inflation with Hubble constant $H_{\rm infl}$, 
the scalar $\tau$ acquired fluctuations $\delta\tau\sim H_{\rm infl}$ independent of those of the inflaton.
Such $\tau$ fluctuations are generically expected unless $m_\tau \gtrsim H_{\rm infl}$, 
in which case the motion of $\tau$ and the resulting baryon asymmetry is reduced with respect to our previous estimates.
Barring this case, baryon iso-curvature fluctuations are estimated as $S_B =\delta n_B/n_B \sim \delta \dot{\tau}/\dot{\tau} \sim H_{\rm infl}/h$ \cite{1605.00670}.
Iso-curvature fluctuations on current cosmological scales are observationally constrained to be $S_B \lesssim S_B^{\rm exp}=4~10^{-5}$~\cite{Planck:2018jri},
implying the bound 
\beq m_\tau \lesssim H_{\rm infl} \lesssim S_B^{\rm exp}  h.\eeq
Combining this bound with eq.\eq{fastdecay} leads to
\beq \label{eq:isoc}
m_\tau \lesssim (S_B^{\rm exp})^{3/2}  \bp \sim 10^{12}\GeV,\qquad 
h \lesssim (S_B^{\rm exp})^{1/2}  \bp\sim 10^{16}\GeV.\eeq

%
%\MD{Models of spontanous baryogenesis are known to generate baryonic isocurvature perturbations. In our scenario baryonic fluctuation $S_B$ can be estimed using solution of eq. (43) in the Hubble-damped regime
%\begin{equation}
%S_B = \frac{\delta n_B(\tau)}{n_B} = \frac{d \ln n_B}{d \tau} \delta \tau = \frac{d \ln \dot{\tau}}{d \tau} \delta \tau = \frac{1}{\tau_f - \tau_i} \frac{H_{\rm inf}}{2\pi h} \gtrsim \frac{m_\tau}{h}
%\end{equation}
% put constraint $S_B\lesssim 4\times10^{-5}$. We immediately notice it also bounds the ratio $T_\Gamma/T_{\rm osc}<S_B$. Assuming modulus energy does not dominate the total energy density, we get an upper constraint on the mass $m_\tau \lesssim 10^{12}$~GeV.}
%

\subsection{Comparison with similar mechanisms}\label{sec:comparison}
Finally, we can now compare the proposed baryogenesis mechanism with
{\em spontaneous baryogenesis}, which arises from a shift-symmetric coupling to the baryon current~\cite{Cohen:1987vi,Cohen:1988kt},
and with axion baryogenesis, which relies on shift-symmetric axion couplings to particle currents. 
These possibilities employ the pseudo-Goldstones of spontaneously broken global U(1) symmetries: baryon-number, or Peccei-Quinn.
As a result, the axion has a small mass $m_a$ protected by a shift symmetry, so axion baryogenesis can only be operative 
at $T \sim \sqrt{m_a\bp}$.
This is comparable to the decoupling temperature of weak sphalerons, $T_{\rm dec}\approx 130\GeV$, 
leading to a too small baryon asymmetry $n_B/n_\gamma\sim T_{\rm dec}/\bp$.

Our scenario avoids this problem because it does not rely on the conservation of any global U(1) charge.
Lepton and baryon numbers can be only approximate accidental symmetries, as in the Standard Model.
Unlike a pseudo-Goldstone boson, the scalar $\tau$ 
can be arbitrarily heavy and features well-motivated CP-violating couplings that break shift symmetry and contribute to baryogenesis.

It is interesting to compare our mechanism to various works in the literature that explored how to avoid a too small baryon asymmetry.
One possibility is assuming non-standard EW or QCD phase transitions~\cite{1407.0030,1811.03294,2006.03148,2412.03902}.
In our context these phase transitions play no role and can remain standard.
Another possibility is assuming a much faster axion evolution $\dot{a}/f \gg m_a$ combined with Peccei-Quinn charge conservation.
This solution however leads to an extra problem: over-production of axion dark matter~\cite{1910.02080}.
This extra problem can be avoided by ``fast lepto-axiogenesis''~\cite{2006.05687}.
Similarly to our scenario, this relies on Majorana neutrino masses.
Unlike our scenario, it also relies on a fast phase rotation and conserved PQ charge.
Ref.~\cite{2006.04809} shows that substituting the axion with an axion-like particle allows to generate
baryogenesis and the dark matter abundance at the weak phase transition,
relying on a PQ charge, on a fast rotation and on a strong enough coupling~\cite{2006.04809}.
These ingredients are not present in our scenario, where couplings can be very suppressed.
Furthermore, unlike our $\tau$ modulus, axion-like particles do not introduce CP-violation in Yukawa couplings.

\section{Conclusions}\label{sec:concl}
We have demonstrated that baryogenesis can arise from the cosmological evolution of a CP-breaking 
modulus scalar $\tau$, either during the big bang or around the end of inflation.
Fundamental Standard Model parameters depend on the vacuum expectation value of $\tau$, 
acquiring  complex CP-violating values.
Section~\ref{sec:U1} presented a simple QFT realisation of this physics, where a local U(1) symmetry
is spontaneously broken by two scalars,
such that their relative phase is the physical CP-breaking scalar $\tau$.
Section~\ref{sec:mod} presented a realisation in which $\tau$ is the modulus associated with $\SL(2,\mathbb{Z})$ invariance,
appearing in the QFT as a remnant of super-string toroidal compactifications that allow for CP-breaking complex parameters.
Both theories are described by the general effective action of eq.\eq{Lagtau}, that contains Yukawa couplings $Y_{{\rm u,d,e},\nu}(\tau)$, 
theta terms $\theta_{2,3}(\tau)$ for $\SU(2)_L$ and $\SU(3)_c$,
and couplings of $\partial_\mu \tau$ to the various particle currents.
We included right-handed neutrinos with Majorana mass terms $M(\tau)$ such that lepton number $L$ is broken,
while $\SU(2)_L$ sphalerons break $B+L$.

\smallskip

The cosmological time variation of $\tau$ --- possibly, though not necessarily, evolving from a CP-conserving point to a CP-violating minimum of its potential ---
induces a time variation of the phases $\theta_Y(\tau)$ of the SM Yukawa couplings $Y$ and of the theta terms $\theta_{2,3}(\tau)$.
Combined with the couplings of $\partial_\mu \tau$ to SM particle currents, these form the rephasing-invariant combinations of eq.s\eq{muYu},\eq{muYe},\eq{muS},
which act as sources for chemical potentials.
Their role in driving baryogenesis is described by the time evolution equations summarized in section~\ref{sec:evo},
that lead to thermal equilibrium approximate solutions discussed in section~\ref{sec:eqsol}.
We emphasise a non-trivial point: naively one could expect that $\dot\theta_Y$ merely source a chiral asymmetry $\mu_{f_L} = - \mu_{f_R}$ in fermions $f$.
The baryon asymmetry emerges when considering the chiral nature of the weak interactions.

\smallskip

The simplest possibility is that $\tau$ evolves in time while $\Delta L=2$ rates, related to neutrino masses, go out of equilibrium at a decoupling
temperature $T_{\rm dec}\sim 10^{11}\GeV$.
In this scenario the measured baryon asymmetry is reproduced as $n_B/n_\gamma\sim T_{\rm dec}/\bp$.
This mechanism is viable if the $\tau$ scalar is relatively light, with a mass around $m_\tau\sim H \sim T^2/\bp \sim 10\TeV$~\cite{Cohen:1987vi,Cohen:1988kt}.
This possibility was discussed in section~\ref{sec:T11}, where a modular realisation was provided as an example.

A heavier $\tau$ is expected, as its mass $m_\tau$ is not protected by any shift symmetry.
Modular baryogenesis can also operate at a higher temperature $T \sim T_{\rm RH}\lesssim 0.003\bp$, with $\tau$ having a mass up to the inflaton mass,
$m_\tau \sim H \sim T_{\rm RH}^2/\bp$,
and potentially reaching the upper limit  $10^{12}$~GeV allowed by baryon iso-curvature perturbations, eq.\eq{isoc}.
This scenario remains viable provided that the excessive baryon asymmetry is partially diluted, either through entropy release from inflaton decays or via $\Delta L = 2$ scatterings related to Majorana neutrino masses.
This possibility was discussed in section~\ref{sec:T>>11}, where a modular example was provided.

The modular baryogenesis mechanism can also be implemented in supersymmetric theories and is compatible
with the supersymmetric modular solution to the strong CP problem~\cite{2305.08908}.

\footnotesize

\paragraph{Acknowledgements.} 
We thank Ferruccio Feruglio and Michael Zantedeschi for comments.
A.T. is funded by the European Union, NextGenerationEU, 
National Recovery and Resilience Plan
(Mission~4, Component~2) 
under the project \textit{MODIPAC: Modular Invariance in Particle Physics and Cosmology} (CUP~C93C24004940006).


\begin{thebibliography}{nnn}\bibitem{2401.15054}
\article[2401.15054]{N. Sch{\" o}neberg}{JCAP}{06}{006}{2024}
{\href{https://doi.org/10.1088/1475-7516/2024/06/006}{The 2024 BBN baryon abundance update}}.


\bibitem{2003.01100}
\article[2003.01100]{L. Di Luzio, M. Giannotti, E. Nardi, L. Visinelli}{Phys.Rept.}{870}{1}{2020}
{\href{https://doi.org/10.1016/j.physrep.2020.06.002}{The landscape of QCD axion models}}.


\bibitem{2305.08908}
\article[2305.08908]{F. Feruglio, A. Strumia, A. Titov}{JHEP}{07}{027}{2023}
{\href{https://doi.org/10.1007/JHEP07(2023)027}{Modular invariance and the QCD angle}}.


\bibitem{2404.08032}
\article[2404.08032]{J.T. Penedo, S.T. Petcov}{JHEP}{10}{172}{2024}
{\href{https://doi.org/10.1007/JHEP10(2024)172}{Finite modular symmetries and the strong CP problem}}.


\bibitem{2406.01689}
\article[2406.01689]{F. Feruglio, M. Parriciatu, A. Strumia, A. Titov}{JHEP}{08}{214}{2024}
{\href{https://doi.org/10.1007/JHEP08(2024)214}{Solving the strong CP problem without axions}}.


\bibitem{1407.0030}
\article[1407.0030]{G. Servant}{Phys.Rev.Lett.}{113}{171803}{2014}
{\href{https://doi.org/10.1103/PhysRevLett.113.171803}{Baryogenesis from Strong $CP$ Violation and the QCD Axion}}.


\bibitem{1811.03294}
\article[1811.03294]{K.S. Jeong, T.H. Jung, C.S. Shin}{Phys.Rev.D}{101}{035009}{2020}
{\href{https://doi.org/10.1103/PhysRevD.101.035009}{Adiabatic electroweak baryogenesis driven by an axionlike particle}}.


\bibitem{1910.02080}
\article[1910.02080]{R.T. Co, K. Harigaya}{Phys.Rev.Lett.}{124}{111602}{2020}
{\href{https://doi.org/10.1103/PhysRevLett.124.111602}{Axiogenesis}}.


\bibitem{2006.04809}
\article[2006.04809]{R.T. Co, L.J. Hall, K. Harigaya}{JHEP}{01}{172}{2021}
{\href{https://doi.org/10.1007/JHEP01(2021)172}{Predictions for Axion Couplings from ALP Cogenesis}}.


\bibitem{2006.05687}
\article[2006.05687]{R.T. Co, N. Fernandez, A. Ghalsasi, L.J. Hall, K. Harigaya}{JHEP}{03}{017}{2021}
{\href{https://doi.org/10.1007/JHEP03(2021)017}{Lepto-Axiogenesis}}.


\bibitem{2503.04888}
\heparticle[2503.04888]{A. Bodas, R.T. Co, A. Ghalsasi, K. Harigaya, L.-T. Wang}{Acoustic Misalignment Mechanism for Axion Dark Matter}.


\bibitem{Fujikawa:1979ay}
\article{K. Fujikawa}{Phys.Rev.Lett.}{42}{1195}{1979}
{\href{https://doi.org/10.1103/PhysRevLett.42.1195}{Path Integral Measure for Gauge Invariant Fermion Theories}}.


\bibitem{2501.16427}
\heparticle[2501.16427]{A. Strumia}{Solving the strong CP problem}.


\bibitem{2406.02527}
\article[2406.02527]{B.-Y. Qu, G.-J. Ding}{JHEP}{08}{136}{2024}
{\href{https://doi.org/10.1007/JHEP08(2024)136}{Non-holomorphic modular flavor symmetry}}.


\bibitem{Cohen:1987vi}
\article{A.G. Cohen, D.B. Kaplan}{Phys.Lett.B}{199}{251}{1987}
{\href{https://doi.org/10.1016/0370-2693(87)91369-4}{Thermodynamic Generation of the Baryon Asymmetry}}.


\bibitem{hep-ph/9911315}
\article[hep-ph/9911315]{R. Barbieri, P. Creminelli, A. Strumia, N. Tetradis}{Nucl.Phys.B}{575}{61}{2000}
{\href{https://doi.org/10.1016/S0550-3213(00)00011-0}{Baryogenesis through leptogenesis}}.


\bibitem{1804.05066}
\article[1804.05066]{K. Moffat, S. Pascoli, S.T. Petcov, H. Schulz, J. Turner}{Phys.Rev.D}{98}{015036}{2018}
{\href{https://doi.org/10.1103/PhysRevD.98.015036}{Three-flavored nonresonant leptogenesis at intermediate scales}}.


\bibitem{2006.03148}
\article[2006.03148]{V. Domcke, Y. Ema, K. Mukaida, M. Yamada}{JHEP}{08}{096}{2020}
{\href{https://doi.org/10.1007/JHEP08(2020)096}{Spontaneous Baryogenesis from Axions with Generic Couplings}}.


\bibitem{hep-ph/0310123}
\article[hep-ph/0310123]{G.F. Giudice, A. Notari, M. Raidal, A. Riotto, A. Strumia}{Nucl.Phys.B}{685}{89}{2004}
{\href{https://doi.org/10.1016/j.nuclphysb.2004.02.019}{Towards a complete theory of thermal leptogenesis in the SM and MSSM}}.


\bibitem{2503.07752}
\heparticle[2503.07752]{F. Capozzi, W. Giar{\` e}, E. Lisi, A. Marrone, A. Melchiorri, A. Palazzo}{Neutrino masses and mixing: Entering the era of subpercent precision}.


\bibitem{hep-ph/0606054}
\heparticle[hep-ph/0606054]{A. Strumia, F. Vissani}{Neutrino masses and mixings and...}.


\bibitem{2201.02020}
\article[2201.02020]{P.P. Novichkov, J.T. Penedo, S.T. Petcov}{JHEP}{03}{149}{2022}
{\href{https://doi.org/10.1007/JHEP03(2022)149}{Modular flavour symmetries and modulus stabilisation}}.


\bibitem{2411.18603}
\heparticle[2411.18603]{G.-J. Ding, S.-Y. Jiang, Y. Xu, W. Zhao}{Modular invariant inflation, reheating and leptogenesis}.


\bibitem{2405.06497}
\article[2405.06497]{G.-J. Ding, S.-Y. Jiang, W. Zhao}{JCAP}{10}{016}{2024}
{\href{https://doi.org/10.1088/1475-7516/2024/10/016}{Modular invariant slow roll inflation}}.


\bibitem{2405.08924}
\article[2405.08924]{S.F. King, X. Wang}{JCAP}{07}{073}{2024}
{\href{https://doi.org/10.1088/1475-7516/2024/07/073}{Modular invariant hilltop inflation}}.


\bibitem{2407.12081}
\heparticle[2407.12081]{G.F. Casas, L.E. Ib{\' a}{\~ n}ez}{Modular Invariant Starobinsky Inflation and the Species Scale}.


\bibitem{Planck:2018vyg}
\article[1807.06209]{{\sc Planck} Collaboration}{Astron.Astrophys.}{641}{A6}{2020}
{\href{https://doi.org/10.1051/0004-6361/201833910}{Planck 2018 results. VI. Cosmological parameters}}.


\bibitem{1805.02656}
\article[1805.02656]{X. Chen, Y. Wang, Z.-Z. Xianyu}{JHEP}{09}{022}{2018}
{\href{https://doi.org/10.1007/JHEP09(2018)022}{Neutrino Signatures in Primordial Non-Gaussianities}}.


\bibitem{1106.2814}
\article[1106.2814]{A. Salvio, P. Lodone, A. Strumia}{JHEP}{08}{116}{2011}
{\href{https://doi.org/10.1007/JHEP08(2011)116}{Towards leptogenesis at NLO: the right-handed neutrino interaction rate}}.


\bibitem{1112.1205}
\article[1112.1205]{M. Laine, Y. Schroder}{JHEP}{02}{068}{2012}
{\href{https://doi.org/10.1007/JHEP02(2012)068}{Thermal right-handed neutrino production rate in the non-relativistic regime}}.

\bibitem{1605.00670}
\article[1605.00670]{A. De Simone, T. Kobayashi}{JCAP}{08}{052}{2016}
{\href{https://doi.org/10.1088/1475-7516/2016/08/052}{Cosmological Aspects of Spontaneous Baryogenesis}}.

\bibitem{Planck:2018jri}
\article[1807.06211]{{\sc Planck} Collaboration}{Astron.Astrophys.}{641}{A10}{2020}
{\href{https://doi.org/10.1051/0004-6361/201833887}{Constraints on inflation}}.


\bibitem{Cohen:1988kt}
\article{A.G. Cohen, D.B. Kaplan}{Nucl.Phys.B}{308}{913}{1988}
{\href{https://doi.org/10.1016/0550-3213(88)90134-4}{Spontaneous baryogenesis}}.


\bibitem{2412.03902}
\heparticle[2412.03902]{W. Chao}{Eogenesis via the High-scale Electroweak Symmetry Restoration}.


\end{thebibliography}
\end{document}